\documentstyle[prl,aps,epsfig,multicol]{revtex}
\newcommand{\be}{\begin{equation}}
\newcommand{\ee}{\end{equation}}

\newcommand{\bea}{\begin{eqnarray}}
\newcommand{\eea}{\end{eqnarray}}
\newcommand{\MeV}{{\rm MeV}}
\newcommand{\eV}{{\rm eV}}

\newcommand{\Be}{{\rm Be}}
\newcommand{\Bo}{{\rm B}}
\newcommand{\Cl}{{\rm Cl}}
\newcommand{\Ga}{{\rm Ga}}
\newcommand{\SK}{{\rm SK}}
\newcommand{\SNO}{{\rm SNO}}
\newcommand{\CC}{{\rm CC}}
\newcommand{\SNOCC}{{\rm SNO-CC}}
\newcommand{\expt}{{\rm expt}}
\begin{document}
\title{Bounds on neutrino magnetic moment tensor from solar
neutrinos }
\author{
Anjan S. Joshipura and  Subhendra Mohanty}

\address{ Physical Research Laboratory,
Navrangpura, Ahmedabad - 380 009, India\\ }

\maketitle

\begin{abstract}
Solar neutrinos with non-zero magnetic moments will contribute to
the electron scattering rates in the Super-Kamiokande experiment.
The magnetic moment scattering events in Super-K can be
accommodated in the standard VO or MSW solutions by a change of
the parameter space of mass square difference and mixing angle-but
the shifted neutrino parameters obtained from Super-K will (for
some values of neutrino magnetic moments) become incompatible with
the fits from SNO, Gallium and Chlorine experiments. We compute
the upper bounds on the Dirac and Majorana magnetic moments of
solar neutrinos by simultaneously fitting all the observed solar
neutrino rates. The bounds the magnetic moment matrix elements are of the
order of $
10^{-10} \mu_B$.

\end{abstract}


\noindent{\bf 1. Introduction}
\\
\\
 Solar neutrino experiments with Chlorine \cite{cl}, Gallium
\cite{ga}, and water cerenkov detectors \cite{sk-sol},\cite{sno},
show  that there is a deficit of $\nu_e$ at the Earth compared to
the predictions \cite{bp2000} of the standard solar model (SSM).
It is commonly accepted that vacuum oscillations (VO) or matter
induced MSW conversions, with large mixing angle (LMA) or small
mixing angle (SMA), can account for the deficit of solar
neutrinos. The available experimental results allow not only a
test of the neutrino oscillation hypothesis but also offer
possibilities of constraining new physics, {\it e.g.} neutrino
magnetic moment \cite{mom1,mom2,mom3}, neutrino decay
\cite{decay}, flavour changing neutral currents, etc.

Neutrinos with non-zero magnetic moments will contribute to the
$\nu e^-$ scattering rates at Super-K. The electron scattering
rate at Super-K (defined as the ratio of average number of
electron scatterings observed  and the theoretical prediction of
$\nu_e e^-$ elastic scattering in SSM) is given by \be R_{SK}={1
\over \langle \sigma_e \rangle}\left( \langle \sigma_e P_{ee}
\rangle + \langle \sigma_{\mu,\tau}(1-P_{ee}) \rangle + \langle
\sigma_{mag} \rangle\right) \label{rsk} \ee Here the first two
terms denote the standard $\nu_e e^-$ and $\nu_{\mu, \tau} e^{-}$
scattering processes respectively , $P_{ee}$ being the $\nu_e$
survival probability. The last term in eq.(\ref{rsk}) corresponds
to the scattering events $\nu_i e^- \rightarrow \nu_j e^-$ due to
non-zero neutrino magnetic moments. Unlike, Super-K and SNO,  the
other solar detectors do not probe the neutral current
interactions and are thus not sensitive to the magnetic moment
scattering. The previous bounds on neutrino magnetic moment from
solar neutrinos have made use of the only Super-K results
\cite{mom1,mom2} or have combined Super-K and SNO results
\cite{mom3} together. The other experiments however indirectly
influence the bound on magnetic moment through a combined
chi-square analysis. We use results from all experiments here
concentrating mainly on rates to put bounds on elements of the
magnetic moment tensor $\mu_{ij}$ defined in the neutrino mass
basis.

The magnetic moment scatterings are helicity flipping processes
and are more appropriately described in the mass basis unlike the
non-flip weak interaction processes which are described in the
flavor basis. The electron neutrinos produced at the Sun arrive as
a linear combination of mass eigenstates at the detector. The
probability amplitude of the $\nu_i$ mass eigenstate at the
detector is denoted by ${\cal A}_i(L)$ and $\nu_i e^-\rightarrow
\nu_j e^- $ scattering amplitude is proportional to $\mu_{ij}$. It
is assumed \cite{mom2} that prior to scattering the neutrinos are
a linear combination of different mass eigenstates, but after
scattering the neutrino beam is a incoherent sum of different mass
states. A field theoretic justification  of this is given in
\cite{grimus}.
 The probability amplitude of the $\nu_i$ mass eigenstate
at the detector is denoted by ${\cal A}_i(L)$ and the $\nu_i
e^-\rightarrow \nu_j e^- $ scattering amplitude is proportional to
$\mu_{ij}$. The magnetic moment differential scattering cross
section at the detector is then given by \be {d\sigma_{mag}\over
dT} = \mu_{eff}^2  \times {\pi \alpha^2 \over m_e^2}({1\over T}
-{1\over E_\nu}) \label{dsmag} \ee  $T$ is the recoil energy of
the electrons, $E_\nu$ the
 initial neutrino energy and the magnetic moments $\mu_{ij}$ are
 in units of Bohr magnetron.
 The effective magnetic moment $\mu_{eff}$ can be written in terms of the
intrinsic neutrino magnetic moments $\mu_{ij}$ as \be \mu_{eff}^2=
{\displaystyle \sum_{j}} |{\displaystyle \sum_{i}}{\cal A}_{iL}
\mu_{ij}|^2 \label{mueff}
 \ee
The magnetic moment tensor $\mu_{ij}$ is antisymmetric if
neutrinos are Majorana particles. For Dirac neutrinos, $\mu_{ij}$
are arbitrary complex numbers but in special case of CP conserving
theory, they are real and symmetric \cite{mom1}. We shall assume
this to be the case.
 The
amplitude ${\cal A}_{iL}$ of the $\nu_i$ mass eigenstate at the
detector depends upon the oscillation model which solves the solar
neutrino problem.

We consider the case of vacuum oscillations, and  matter induced
MSW conversions in turn and derive bounds on $\mu_{ij}$ for each
of these cases separately.
\\
\\
\noindent {\bf 2. Vacuum Oscillations}
\\
\\
 Neutrinos
are produced in the solar core as the $\nu_e$ flavor eigenstate,
and the amplitude for detecting them at a distance $L$ as a mass
eigenstate is given by \be {\cal A}_{i}(L)= U_{e i} e^{i E_i L}
~.\ee The elements of mixing matrix $U$ are constrained by
atmospheric neutrino and CHOOZ experiments to be of the following
form:
 \bea
\pmatrix{
  \nu_e \cr
  \nu_\mu\cr
  \nu_\tau}
= \pmatrix{
  c_1& s_1 & 0 \cr
  -c_2 s_1 & c_2 c_1 & s_2\cr
  s_2 s_1& -s_2 c_1& c_2}
 ~~\pmatrix{
  \nu_1 \cr
  \nu_2\cr
  \nu_3}~.
 \label{u}
\eea

We have approximated $U_{e3} \sim 0$ in order to account for the
negative results obtained in  $\nu_e$ disappearance experiment at
CHOOZ \cite{chooz}.  The small value of $U_{e3}$ (taken here as
zero) also prevents the mixing of the third neutrino in vacuum. In
addition to the above mixing one also needs $\delta m_{12}^2=
m_2^2-m_1^2\approx 10^{-4}-10^{-11}$ eV$^2$ and $\delta
m_{23}^2\equiv m_3^2-m_2^2\approx 10^{-3}$ eV$^2$ to account for
the solar and atmospheric \cite{sk-atm} neutrino problems
respectively. The required value of $\delta m_{23}^2$ is much
larger than the value of the effective mass square
$2\sqrt{2}G_FEN_e$ of the electron neutrino at the solar core.
This suppresses mixing of the third neutrino in matter as well
\cite{kras} which decouples from the rest in case of the MSW
solution to the solar neutrino problem.

Using the mixing matrix (\ref{u}) the $\nu_e $ survival
probability in vacuum from Sun to Earth is given by the
expressions \bea
 P_{ee} &=& c_1^4 + s_1^4~ + 2 (c_1 s_1)^2~ \cos({\delta m_{12}^2 L \over 2
 E_{\nu}})~ \eea
where  $L=1.5 \times 10^{13}\, {\rm cm}\, =500.3$ s is the
Earth-Sun distance.

Using the mixing matrix elements from eq.(\ref{u}), we find that
the expression for $\mu_{eff}^2$ for VO is given by \bea
(\mu_{eff}^2)_{VO} &=&{\displaystyle \sum_{j}}
\left(|{\displaystyle \sum_{i}} U_{e i} e^{i E_i L} \mu_{ij}|^2
\right) \nonumber \\ &=& c_1^2 \left(\mu_{11}^2 + \mu_{12}^2 +
\mu_{13}^2 \right) + s_1^2 \left(\mu_{21}^2 + \mu_{22}^2
+\mu_{23}^2 \right) \nonumber \\
 &+& 2 c_1 s_1 \left( \mu_{11} \mu_{21} +\mu_{12} \mu_{22} +
\mu_{13} \mu_{23} \right) \cos ( {\delta m_{12}^2 L \over 2 E_\nu}
) \label{mueffvo}
 \eea
 This  is the three generation generalization of the
 expression  for $(\mu_{eff}^2)_{VO} $ first derived by Beacom and
 Vogel \cite{mom2}.
The $\nu_3$ state can be produced after the electron scattering by
$\nu_1$ and $\nu_2$ in the solar neutrino beam, therefore the
transition moments $\mu_{13}$ and $\mu_{23}$ appear in the
expression above. However since the solar neutrino beam does not
contain $\nu_3$, there is no $\mu_{3i}$ dependence in $\mu_{eff}$
and $\mu_{33}$ cannot be constrained by solar neutrino
experiments.

Vacuum solution requires $\delta m_{12}^2\sim 10^{-10}$eV$^2$. The
oscillatory term in eq.(\ref{mueffvo}) does not get averaged out
in this case. This term by itself is not positive definite and
prevents obtaining bounds on all the moments appearing in
eq.(\ref{mueffvo}). We thus need to make simplifying assumptions.
For the case of Dirac neutrinos we assume (as in the Particle Data
Book) that the dominant non-zero elements are diagonal $\mu_{ij}=
\delta_{ij} \mu_i$. With this assumption, eq.(\ref{mueffvo})
simplifies to the form \be \mu_{eff}^2= c_1^2 \mu_1^2 + s_1^2
\mu_2^2. \label{diracvo}\ee Since each term \ref{diracvo} is
positive definite we can put upper bounds on $\mu_1^2$ and
$\mu_2^2$ by taking them one at a time.

For the case of Majorana neutrinos the magnetic moment matrix
$\mu_{ij}$ is anti-symmetric. We set $\mu_{11}=\mu_{22}=0$ and
find that $\mu_{eff}$ can be given an upper and lower bounds
\be
\mu_{12}^2 +(c_1 |\mu_{13}| - s_1 |\mu_{23}| )^2 \leq \mu_{eff}^2
\leq \mu_{12}^2 +(c_1 |\mu_{13}| + s_1 |\mu_{23}| )^2.
\label{mueff1}
 \ee From this we see that
$\mu_{eff}^2$ can be written as a sum of two positive definite
quantities and therefore one can take each term separately in
order to put an upper  bound on $\mu_{12}$. To put bounds on
$\mu_{13}$ and $\mu_{23}$ we need to further assume that there is
no cancellation between the two terms in the the expression $(c_1
|\mu_{13}| - s_1 |\mu_{23}| )^2$ i.e  either $|\mu_{23}|$ or $
|\mu_{13}|$ is much larger than the other.
\\
\\
\noindent{\bf 3. MSW conversion}
\\
\\
 If the flavor conversion of neutrinos in the Sun is by MSW
mechanism then we have  different expressions for $\mu_{eff}$. As
discussed already, only two energy eigenstates $\nu_1$ and $\nu_2$
corresponding to the lighter neutrinos participate in the MSW
conversion. In the core of the Sun $E_1
> E_2$, there is a level crossing at the resonance point after
which $E_2 > E_1$ {\it i.e.} $m_2 > m_1 $ in vacuum. The
probability of $\nu_e \rightarrow \nu_1$ just after level crossing
is
\be
P_1  =P_J s_m^2 + (1-P_J) c_m^2
\label{p1}
 \ee
 Here, the first
term stands for $\nu_e$ going to $\nu_2$ by mixing with
probability $s_m^2$ and then jumping to $\nu_1$ with probability
$P_J$ at the level crossing. The second term means that $\nu_e$
goes to $\nu_1$ by mixing and then does not jump  to $\nu_2$ with
probability $(1-P_J)$ at the level crossing. The probability of
$\nu_e \rightarrow \nu_2$ just after level crossing is \be P_2
=(1-P_1)=(1-P_J) s_m^2 + P_J c_m^2. \ee In the formulas above, the
Landau-Zener jump probability is given by \bea P_{J} &=& {\exp(-b
s_1^2 /E_\nu)-\exp(-b /E_\nu)\over 1-\exp(-b /E_\nu)} \label{lz}\\
b&=&  {\pi \over 4}\, \left({\delta m^2   \over
 |\dot A /A|_{res}} \right)
\simeq 10^9\, \left({\delta m^2 \over \eV^2}\right)\, \MeV
\eea
 and $A= 2 \sqrt{2} E_\nu G_F N_e$.
The mixing angle in matter in the Sun is given in terms of the
vacuum mixing angle by the expression \bea \cos 2\theta_m &=&
{(-1+\eta(1-2 s_1^2))\over (1- 2 \eta(1-2 s_1^2) +\eta^2)^{1/2}} \\
\eta&=& { \delta m^2 \over A} =6.6\times 10^{-5} {b\over E_\nu}
\eea

We can now use the matrix eq.(\ref{u}) to find the survival
probability of $\nu_e$ at Earth, \be P_{ee}= c_1^2 P_1 + s_1^2 P_2
\ee

Unlike in the vacuum case, ${2 E_\nu\over \delta m_{12}^2}$ is
much smaller than the Sun-Earth distance in case of  the MSW
conversion. Thus phases acquired in oscillations average out
\cite{dighe}. We can thus drop
 the interference terms of the type ${\cal A}_1(L){\cal
 A}_2(L)^*$ in the expression for $\mu_{eff}$ in eq.(\ref{mueff}).
 Substituting  for $|{\cal A}_1(L)|^2= P_1$
and $|{\cal A}_2(L)|^2= P_2$ and we then get \bea
(\mu_{eff}^2)_{MSW} &=&{\displaystyle \sum_{j}}
\left({\displaystyle \sum_{i}} P_i \mu_{ij}^2 \right) \nonumber
\\ &=& P_1 \left(\mu_{11}^2 + \mu_{12}^2 + \mu_{13}^2 \right) +
P_2 \left(\mu_{21}^2 + \mu_{22}^2 +\mu_{23}^2 \right) \nonumber\\
&=& \mu_{12}^2 + P_1 (\mu_{11}^2 + \mu_{13}^2) + (1-P_1)
(\mu_{22}^2 + \mu_{23}^2 ) \label{mueffmsw}
 \eea
The  expression for $(\mu_{eff}^2)_{MSW}$ was first derived for
the two generations case in \cite{mom2}.
 Since, $(\mu_{eff}^2)_{MSW}$ contains a sum of
positive definite terms we can put upper bounds on each term taken
one at a time, using the expression eq.(\ref{p1}) for $P_1$. In
this case, we are able to constrain all $\mu_{ij}$ except
$\mu_{33}$.
\\
\\
\noindent{\bf 4. Experimental rates and bounds on magnetic
moments}
\\
\\
 The electron scattering reaction in Super-K can be written as
\bea R_{\SK} = {\int dE_\nu \sigma_{\nu_e} \Phi_\Bo P_{ee} +\int
dE_\nu \sigma_{\nu_\mu} \Phi_\Bo (1-P_{ee})+\int dE_\nu
\sigma_{mag}(E_\nu) \Phi_\Bo \over \int dE_\nu \sigma_{\nu_e}
\Phi_\Bo } \label{tsk} \eea

The magnetic moment scattering cross section $\sigma_{mag}(E_\nu)$
is obtained by integrating (w.r.t the electron recoil energy $T$
)the differential cross section \ref{dsmag} folded with a detector
response function taken from \cite{fogli}. The $^8 B$ $\nu_e$ flux
is taken from \cite{bp2000} and is given below. Only the Super-K
electron elastic scattering rate depends on the neutrino magnetic
moments.

 The charge-current deuterium dissociation  reaction rate at  SNO
 normalized to its SSM value is given by
\bea R_{\SNO}^{\CC}= {\int dE_\nu \sigma_{\CC} \Phi_\Bo P_{ee}
\over \int dE_\nu \sigma_{\CC} \Phi_\Bo } \label{tsno} \eea The
$\nu_e e^-$ and $\nu_{\mu,\tau} e^-$ elastic scattering cross
section after folding with the detector response function are
tabulated in ref.\cite{Ba-SK}. The deuterium dissociation cross
section is taken from \cite{Ba-snocc}.

The rates of neutrino capture in the Chlorine and Gallium
experiments can be written as \be
 R_\alpha ={ {\displaystyle \sum_{i=pp,\Be,\Bo}}
\int dE_\nu \Phi_i  \sigma_{ \alpha} P_{ee}\over
{\displaystyle \sum_{i=pp,\Be,\Bo}} \int dE_\nu \Phi_i  \sigma_{ \alpha}}
 \ee
where the subscript $\alpha=$ Ga, Cl denotes the experiment and
$i=pp$, $^7$Be, $^8$B denotes the type of neutrino flux from the Sun.
The spectra of the $pp $ and $^8$B neutrinos can be fitted with
the analytical functions,
\bea
\Phi_{pp} &=&(5.95\times 10^{10})
[193.9 (0.931-E_\nu)((0.931-E_\nu)^2 -0.261)^{1/2} E_\nu^2]
\nonumber \\
 \Phi_{\Bo}&=&(5.05 \times 10^{6} )[8.52
\times 10^{-6}(15.1-E_\nu)^{2.75} E_\nu^2 ]  \nonumber
\\
\Phi_{\Be} &=&(4.77 \times 10^{9}) [ \delta(E_\nu -0.862)]
\label{flux}
 \eea
where the neutrino fluxes are in units of cm$^{-2}$s$^{-1}$ and
$E_\nu$ is
in MeV. The first brackets in (\ref{flux}) give the
total flux of neutrinos from the $pp$,Be, and B reactions
and are taken from
BP2000 \cite{bp2000}, and the square brackets give the  spectral
shape \cite{bu}.

The Ga experiments can detect all three types of neutrino fluxes and the
neutrino absorbtion cross section of $Ga$ is given in
\cite{Ba-Ga}. The Chlorine experiment threshold is higher (0.8 MeV) and it
detects only the Be and B neutrinos; we take the absorption cross
section with the tables from ref.\cite{Ba-Cl}.

Using the flux spectrum in equation (\ref{flux}) and the cross
sections \cite{Ba-Ga,Ba-Cl,Ba-SK,Ba-snocc} we can calculate the
theoretical rates for MSW or VO conversion probabilities as a
function of the three unknown parameters: the magnetic moment
matrix elements $\mu_{ij}$, $\delta m_{12}^2$ and the vacuum mixing
angle $\theta_1$. The experimental rates $R_\alpha^{\expt}$ with
one-sigma combined (statistical and systematic) experimental
errors $\Delta_\alpha$ are as follows \cite{cl,ga,sk-sol,sno}:
 \bea
 R_{\Cl}^\expt &=& 0.335 \pm 0.029 \nonumber\\
 R_{\Ga}^\expt &=& 0.584 \pm 0.039 \nonumber\\
 R_{\SK}^\expt &=& 0.459 \pm 0.017 \nonumber\\
 R_{\SNOCC}^\expt &=&0.347 \pm 0.029
 \label{rexp}
\eea
From the theoretical $R_\alpha$, and the experimental
$R_\alpha^{\expt}$,
we compute the total $\chi^2$ for all experiments,
defined as
\be
\chi^2 \equiv {\displaystyle \sum_{\alpha=
\Cl,\Ga,\SK,\SNO}}{(R_\alpha -R_\alpha^{\expt} )^2 \over
\Delta_\alpha^2}. \ee

Setting the parameter $\mu_{eff} =0$ we reproduce the standard
contours of VO (Fig.~ 1.) and MSW (Fig.~ 2)  solutions shown as
continues curves. For $\mu_{eff}=0$ the global minimum of $\chi^2$
occurs in the vacuum region with $\chi_{min}^2\sim 0.3$. We then
take a non-zero $\mu_{eff}$ and plot the regions (shown as dashed
curves) in $\delta m_{12}^2-\tan^2\theta$ plane  corresponding to
$\chi^2_{min} +4.6$ (90\% CL) both for VO (Fig.1) and MSW ( Fig.
2) solutions. These figures correspond to a constant value for
$\mu_{eff}$. It is found that the allowed regions shrinks as we
increase the value of $\mu_{eff}$ and it disappears for some
specific $\mu_{eff}$ which is taken to be the 90\% CL bound on
$\mu_{eff}$. We consider different cases following this procedure.

In case of VO and Dirac neutrino, we choose $\mu_{eff}^2$ equal to
$\mu_1^2 c_1^2$ and $\mu_2^2 s_1^2$ (see eq.(\ref{diracvo}))
respectively. The bounds on $\mu_{1,2}$ obtained using the above
procedure are given by (at $90 \% C.L.$)
 \bea
 \mu_1 &<& 4.5 \times 10^{-10} \mu_B \nonumber\\
 \mu_2 &<& 7.1 \times 10^{-10} \mu_B
 \label{vo1}
 \eea
In case of the Majorana neutrinos we use eq.(\ref{mueffvo}) and
substitute  $\mu_{eff}^2$ by $\mu_{12}^2$ , $c_1^2 |\mu_{13}|^2$
or $s_1^2 |\mu_{23}|^2$ leading to the 90\% CL bounds
\bea |\mu_{12}|&<& 3.7 \times 10^{-10} \mu_B \nonumber ~,\\
|\mu_{13}| &<& 4.5 \times 10^{-10} \mu_B ~~~(~if~~
|\mu_{23}|<<|\mu_{13}|) \nonumber\\ |\mu_{23}| &<& 7.1 \times
10^{-10} \mu_B ~~~ (~if~~ |\mu_{13}| << |\mu_{23}| )
 \label{vo2}
 \eea
Next we take up the MSW solution. We find that the LMA and SMA
allowed parameter space disappears for $\mu_{eff} = 3.7 $ which
translates to the bound on $\mu_{12}$. Next we take $\mu_{eff}
=(1- P_1) \mu^2$ and find that the parameter space vanishes when
$\mu_{eff}=4.3$. This translates to the bounds on
$|\mu_{22}|,\mu_{23}|$. Finally we take $\mu_{eff} = P_1 \mu^2$
and find that the SMA region disappears for $\mu_{eff} =7.8$ and
the LMA region disappears at $\mu_{eff}=20.5$ which translates
into the bounds on $|\mu_{11}|,\mu_{13}|$. These bounds are
summarized below:
 \bea |\mu_{12}| &<& 3.8 \times 10^{-10} \mu_B \nonumber\\
 |\mu_{22}|, |\mu_{23}| &<& 4.3 \times 10^{-10} \mu_B \nonumber\\
 |\mu_{11}|, |\mu_{13}| &<& 7.8 \times 10^{-10} \mu_B ~~(SMA)~~ ;~~20.5 \times 10^{-10} \mu_B ~~(LMA) )
\label{msw1}
 \eea
In (\ref{msw1}) , the bounds on the diagonal elements of
$\mu_{ij}$ are relevant for Dirac neutrinos only and the bounds on
the off-diagonal elements apply to both Dirac and Majorana
neutrinos. The bound on $|\mu_{11}|, |\mu_{13}|$ obtained in case
of the LMA solution is weakest of all bounds. This is due to the
fact that these quantities are multiplied by $P_1$ in
eq.(\ref{mueffmsw}) which is very small for adiabatic transition
occurring in the LMA region.

While the neutrino magnetic moments affect the rates of only the
Super-K experiment, a combined analysis of all experiments was
essential in establishing the bounds given in
eqs.(\ref{vo1}-\ref{msw1}). This can be seen explicitly in Fig.~3.
and Fig.~4 for the VO case and in Fig. 5. and Fig. 6. for the MSW
case. In Fig. 3. we plot the $1.64 \sigma$ allowed regions for
$R_{SK}$ (continuous line) and $R_{SNO}$ (dotted line)  by setting
$\mu_{eff}=0$. We see a large overlap between the SK and SNO
allowed regions for $\mu_{eff}=0$. Next we repeat the same plots
by taking $\mu_{eff}=3.8$. We see that the SK allowed parameter
shifts so that there is no more any overlap with the SNO allowed
region. Therefore a combination of SNO and SK result is crucial in
disallowing $\mu_{eff} > 3.8$.

Similarly in Fig. 5. and Fig. 6. we plot the $1.64 \sigma$ allowed
regions for $\mu_{eff}=0$ and $\mu_{eff}=4.0$ respectively. In
Fig.5. we see overlap between SK allowed regions (between
continuous curves), SNO (between dashed curves) and Gallium
(between dotted curves) in the SMA and the LMA regions. In Fig. 6.
we see that when $\mu_{eff}= 4.0$ the SK allowed parameter space
shifts so that there is no common allowed region for the three
experiments thereby ruling out values of $\mu_{eff} > 4.0$.

In summary, we have used solar neutrino rates to constrain
possible magnetic moment coupling of neutrinos to photons. The
bounds depend upon specific solution of the solar neutrino
problem. If MSW mechanism is mainly responsible for the solar
neutrino conversion then it is possible to constrain all elements
of magnetic moment operator $\mu_{ij}$ except for $\mu_{33}$. For
the VO solution one can obtain bounds (eq.(\ref{vo1})) on two of
the diagonal moments assuming a diagonal $\mu_{ij}$ in case of the
Dirac neutrinos. For Majorana neutrinos, we obtain bounds on
$\mu_{12}$ , $\mu_{13}$ and $\mu_{23}$ given in eq.(\ref{vo2}).

\newpage

\begin{figure}
 \label{Fig.1}
\centering
\includegraphics[width=10cm]{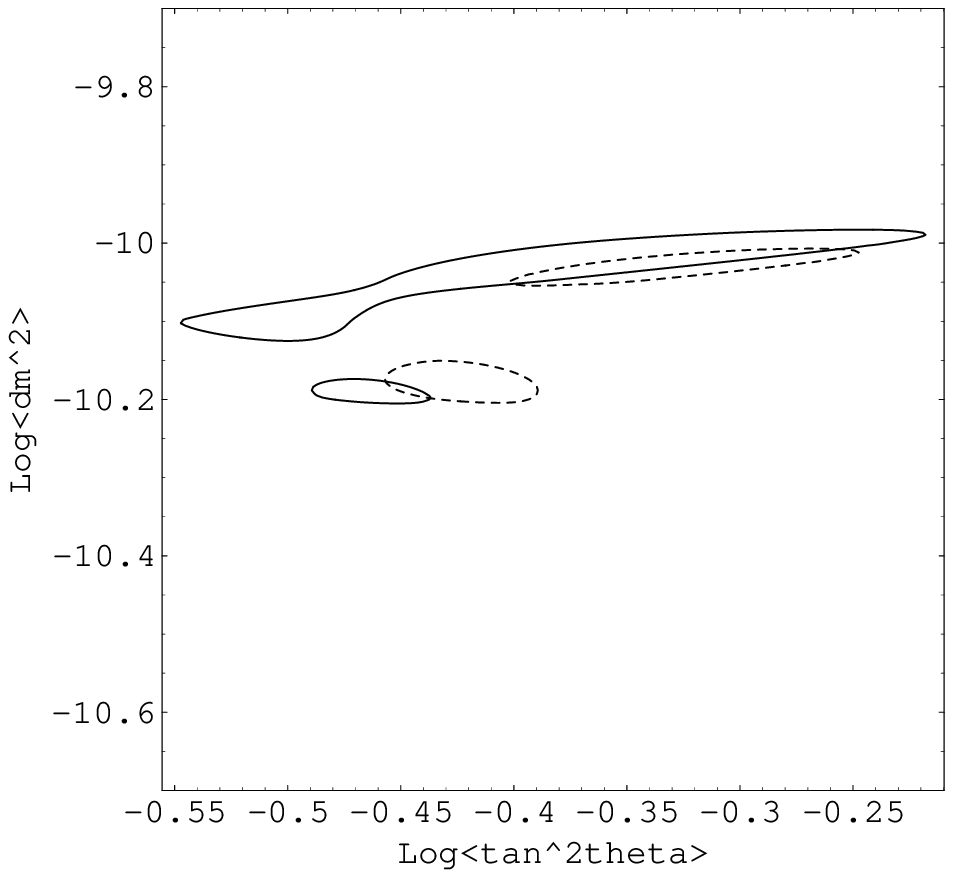}
\caption{ The $90 \%$ allowed parameter space for VO , with
$\mu_{eff}=0$ (enclosed by continuous curve)  and for
$\mu_{eff}=3$ (enclosed by dashed curve).}
\end{figure}

\begin{figure}
 \label{Fig.2}
\begin{center}
\includegraphics[width=10cm]{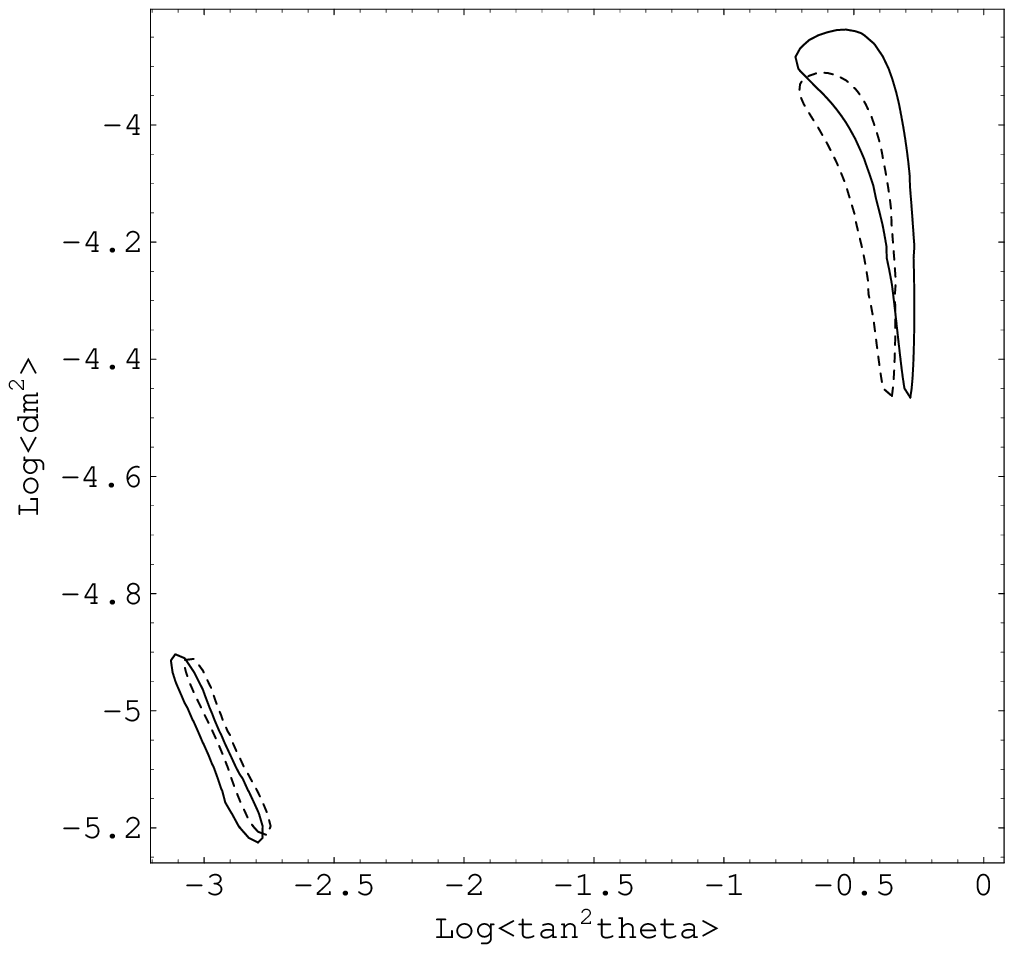}
\end{center}
\caption{ The $90 \%$ allowed parameter space for MSW conversion ,
with $\mu_{eff}=0$ (enclosed by continuous curve)  and for
$\mu_{eff}=3$ (enclosed by dashed curve).   }
\end{figure}

\newpage

\begin{figure}
 \label{Fig.3}
\begin{center}
\includegraphics[width=10cm]{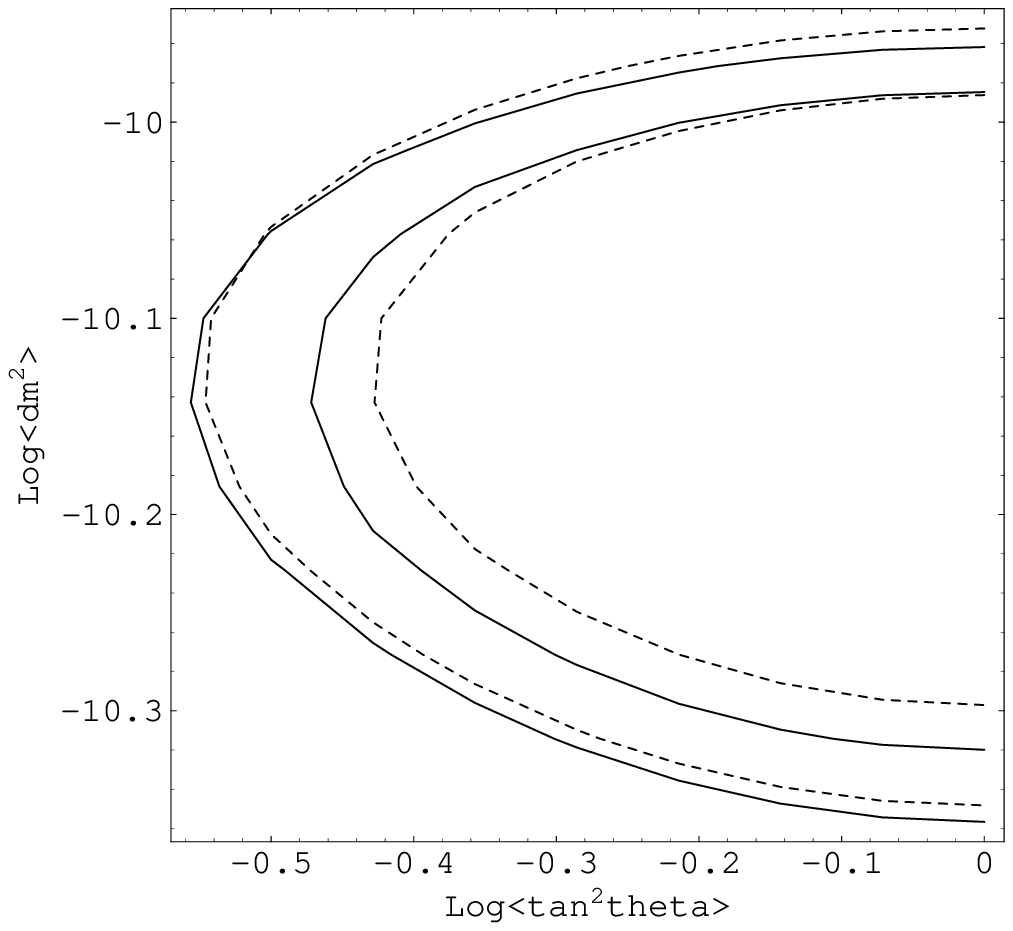}
\end{center}
\caption{The $1.64 \sigma$ VO allowed regions for $R_{SNO}$ (
enclosed by  dashed curves) and $R_{SK}$ ( enclosed by continuous
curves) for $\mu_{eff}=0$. }
\end{figure}

\begin{figure}
 \label{Fig.4}
\begin{center}
\includegraphics[width=10cm]{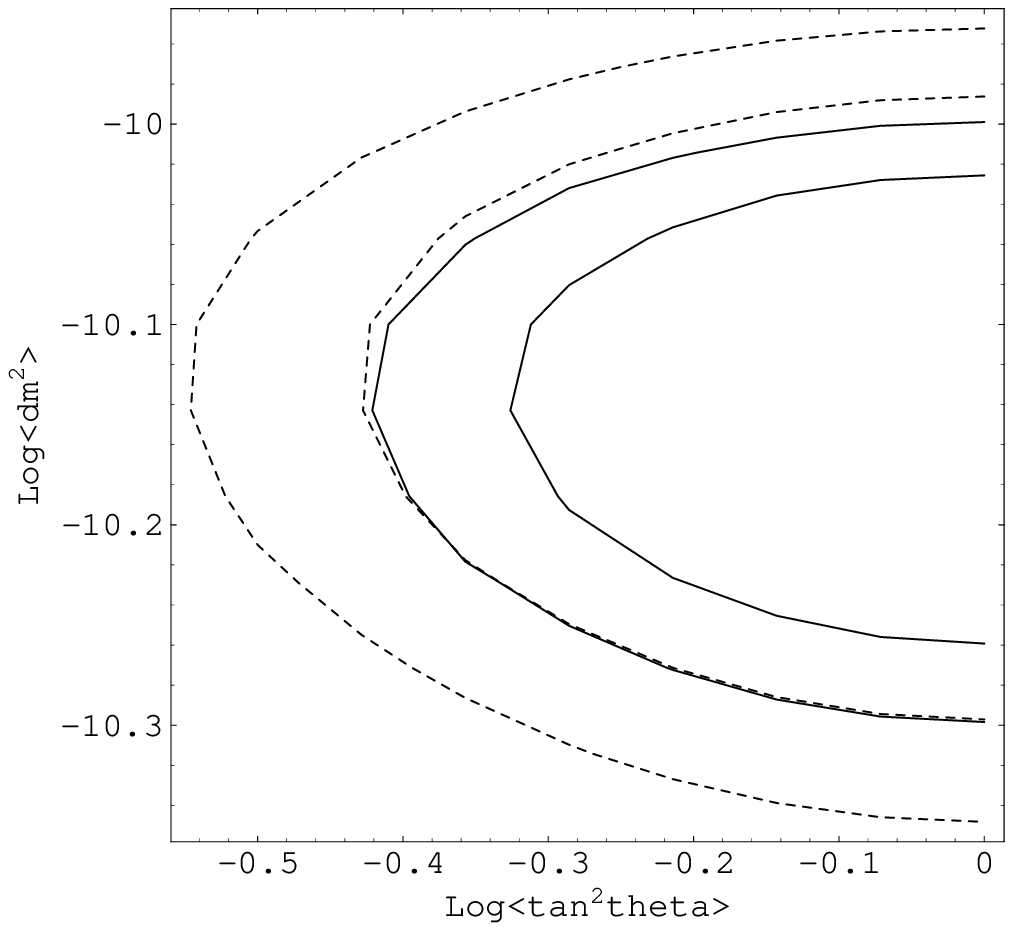}
\end{center}
\caption{The $1.64 \sigma$ VO allowed regions for $R_{SNO}$ (
enclosed by  dashed curves) and $R_{SK}$ ( enclosed by continuous
curves) for $\mu_{eff}=3.8$.}
\end{figure}
\newpage

\begin{figure}
 \label{Fig.5}
\begin{center}
\includegraphics[width=10cm]{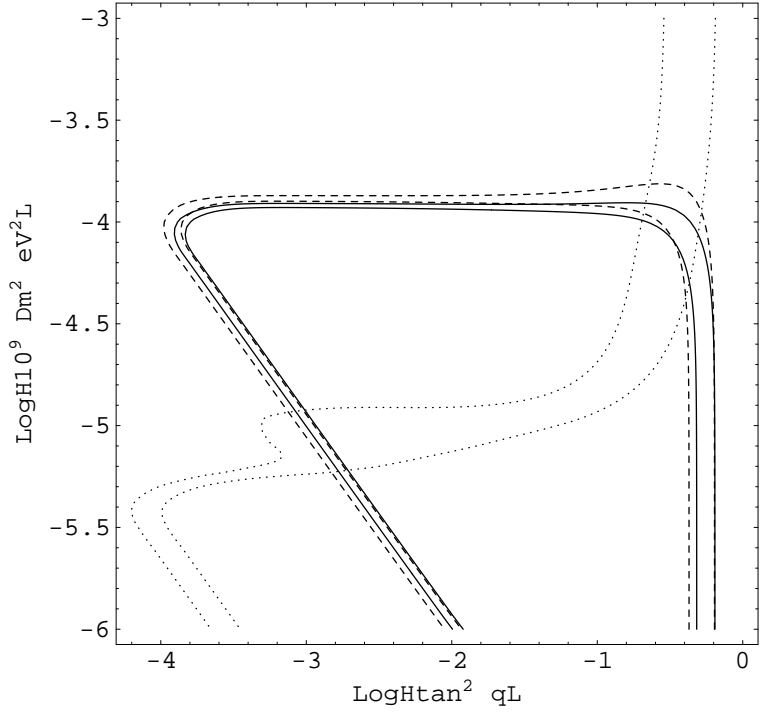}
\end{center}
\caption{The $1.64 \sigma$ MSW allowed regions for $R_{SNO}$ (
enclosed by  dashed curves) and $R_{SK}$ ( enclosed by continuous
curves) and $R_{Ga}$ (enclosed between dotted curves)  for
$\mu_{eff}=0$. There is ovelap in the SMA and LMA regions.}
\end{figure}

\begin{figure}
 \label{Fig.6}
\begin{center}
\includegraphics[width=10cm]{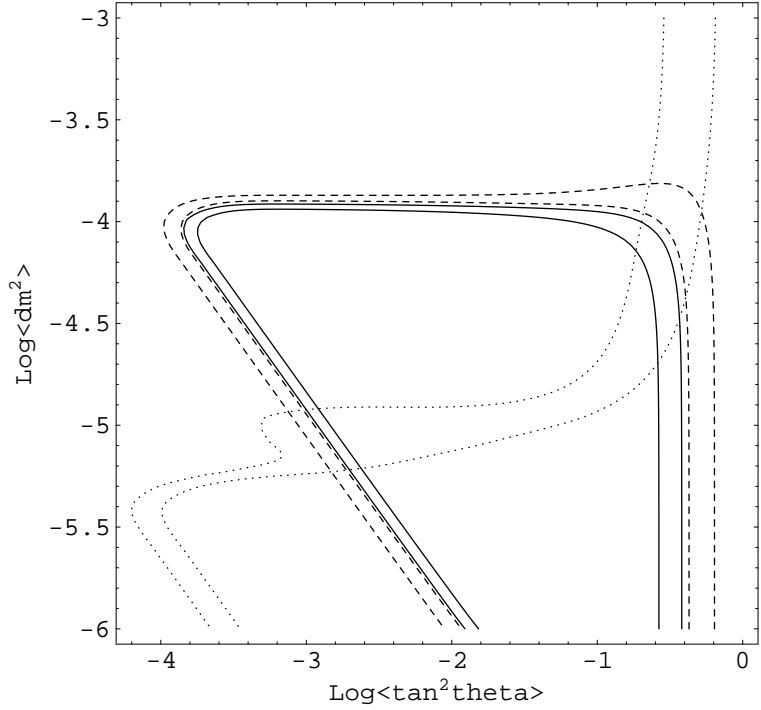}
\end{center}
\caption{The $1.64 \sigma$ MSW allowed regions for $R_{SNO}$ (
enclosed by  dashed curves) and $R_{SK}$ ( enclosed by continuous
curves) and $R_{Ga}$ (enclosed between dotted curves)  for
$\mu_{eff}=4$. The SK allowed region shifts such that there is no
ovelap between the three.}
\end{figure}
\end{document}